\documentclass[aps,prd,reprint,nofootinbib]{revtex4-1}

\usepackage[utf8]{inputenc}
\usepackage{amsmath,amssymb,slashed}
\usepackage{dsfont}
\usepackage{graphicx}
\usepackage{epstopdf}  
\usepackage{slashed}
\usepackage{hyperref}

\usepackage{float}
\usepackage{subcaption}
\usepackage{tikz-feynman}

\begin{document}

\title{Low-energy axial-vector transitions from decuplet to octet baryons}
\author{M\aa ns Holmberg}
\author{Stefan Leupold}

\affiliation{Institutionen f\"or fysik och astronomi, Uppsala Universitet, Box 516, S-75120 Uppsala, Sweden}
\date{\today}

\begin{abstract}
Axial-vector transitions of decuplet to octet baryons are parametrized at low energies guided by a complete and minimal 
chiral Lagrangian up to next-to-leading order. It is pointed out that beyond the well-known leading-order term, there is only 
one contribution at next-to-leading order. This contribution is flavor symmetric. Therefore the corresponding 
low-energy constant can be determined in any strangeness sector. As functions of this low-energy constant, we calculate the 
decay widths and Dalitz distributions for the decays of decuplet baryons to octet baryons, pions, and photons and for the 
weak decay of the Omega baryon to a cascade baryon, an electron, and an anti-neutrino. 
\end{abstract}
%

\maketitle

\section{Introduction and summary}
\label{sec:intro}

One of the interesting processes of neutrino-nucleon scattering is the production of the lowest-lying
Delta state \cite{Alvarez-Ruso:2017oui,Procura:2008ze,Geng:2008bm,Mosel:2015tja,Unal:2018ruo}. 
On the theory side, the relevant quantities are the vector and axial-vector form factors for the transition 
of a nucleon to a Delta. While the vector transition form factors can also be addressed in the electro-production of a Delta on 
a nucleon, information about the axial-vector transition form factors is scarce. Of course, the situation is even 
worse in the strangeness sector. At least, the decay width of the reaction $\Omega^- \to \Xi^0 \, e^- \bar\nu_e$ has been 
measured \cite{Bourquin:1984gd,pdg}. 

The purpose of the present work is to study the low-energy limit of these axial-vector transition form factors and to point out
various ways how to measure these quantities. 
Recently we have established a complete and minimal relativistic chiral Lagrangian of next-to-leading order (NLO) 
for the three-flavor sector including the lowest lying baryon octet and decuplet states \cite{Holmberg:2018dtv}. Based on this
Lagrangian, we will calculate various observables that depend on the low-energy constants (LECs) that enter the axial-vector 
transition form factors.

Dealing with a chiral baryon Lagrangian at NLO means that we carry out tree-level calculations. 
But why do we bother about such NLO tree-level results, if the state of the art seems to be
one-loop calculations \cite{Procura:2008ze,Geng:2008bm}? 
Though axial-vector transition form factors have been calculated at the one-loop level of chiral perturbation theory ($\chi$PT), 
the numerically fairly unknown tree-level contributions have often been formulated based on non-minimal Lagrangians. 
In this way, one assumes a too large number of to be fitted parameters and one misses cross relations between form factors 
and also between different processes. Therefore we have decided to go one step back and analyze the NLO structure, i.e.\ 
tree-level structure of the axial-vector transition form factors. Of course, this can only be the first step of a more detailed 
investigation that should go beyond the NLO level. 

Let us start with the introduction of pertinent axial-vector transition form factors where we can already point out some 
short-comings of previous works. With $B$ denoting a spin-1/2 baryon from the nucleon octet and $B^*$ a spin-3/2 baryon from 
the Delta decuplet, the axial-vector transitions can be written as
\begin{eqnarray}
  \label{eq:axialTFFs}
  \langle B^*(p') \vert j^\mu_A \vert B(p) \rangle = \bar u_\nu(p') \, \Gamma^{\mu\nu} \, u(p)
\end{eqnarray}
with $q=p'-p$ and 
\begin{eqnarray}
  \label{eq:defGamma}
  &&\Gamma^{\mu\nu} = q^\mu q^\nu H_0(q^2) + g^{\mu\nu} H_1(q^2) \nonumber \\
  &&{}+ (\gamma^\mu q^\nu - \slashed q \, g^{\mu\nu}) H_2(q^2) 
  + i \sigma^{\mu\alpha} q_\alpha q^\nu H_3(q^2) \,.
\end{eqnarray}
The advantage of the decomposition \eqref{eq:axialTFFs}, \eqref{eq:defGamma} lies in the fact that contributions to 
the axial-vector transition form factors $H_i$ start 
only at the $i$th chiral order for $i=1,2,3$. This is easy to see because the momentum $q$ carried by the axial-vector current 
is a small quantity of chiral order 1. In particular, this implies that $H_3$ does not receive contributions from 
leading order (LO) and from NLO. The ``regular'' contributions to $H_0$ start also at third chiral order, but $H_0$ receives 
a contribution at LO from a Goldstone-boson pole term. 

For the following rewriting, it is of advantage to recall the equations of motion for the spin-1/2 spinors $u$ and the spin-3/2
Rarita-Schwinger vector-spinors \cite{Rarita:1941mf} $u_\nu$:
\begin{eqnarray}
  \label{eq:eomspinors}
  (\slashed p - m_8) \, u(p) = 0 \,, \quad (\slashed p' - m_{10}) \, u_\nu(p') = 0 
\end{eqnarray}
where $m_{8/10}$ denotes the mass of the considered baryon octet/decuplet state. Note that the mass difference $m_{10}-m_8$ is 
counted as a small quantity. 

The use of the form factor $H_3$ is not very common. Instead of $i \sigma_{\mu\alpha} q^\alpha q_\nu$ one often uses a 
structure \\ $p'_\mu q_\nu - p' \cdot q \, g_{\mu\nu}$; see e.g.\ \cite{Procura:2008ze,Geng:2008bm} and references therein.
The problem with the latter
structure is, however, that it looks like an additional fourth independent term that contributes at NLO. Instead, 
the decomposition \eqref{eq:defGamma} shows that there are only three structures up to and including NLO --- and $H_0$ receives only a 
Goldstone-boson pole contribution, i.e.\ there are only two {\em independent} terms up to and including NLO. 
The structure $p'_\mu q_\nu - p' \cdot q \, g_{\mu\nu}$ is {\em not} required up to and including NLO. 
To make contact between different ways how to parametrize the transition form factors, one can use a Gordon-type identity,
\begin{eqnarray}
  \label{eq:gordon}
  \bar u_\nu \left( -2 p'_\mu + (m_{10}+m_8)\gamma_\mu + q_\mu - i \sigma_{\mu\alpha} q^\alpha \right) u  = 0  \,,
\end{eqnarray}
which can easily be established using the equations of motion \eqref{eq:eomspinors}. 

We can now focus on the three form factors $H_0$, $H_1$ and $H_2$, which receive contributions already at LO or 
NLO, respectively. Using the Lagrangians of LO and NLO from \cite{Holmberg:2018dtv} one obtains 
\begin{eqnarray}
  H_0(q^2) &=& - \frac{h_A}{\sqrt{2}} \, \frac{1}{q^2-m_{\rm GB}^2} \, k_f \,, \nonumber \\
  H_1 &=& \frac{h_A}{\sqrt{2}} \, k_f \,, \nonumber \\
  H_2 &=& -2 c_E \, k_f 
  \label{eq:resHi}
\end{eqnarray}
with the LO low-energy constant $h_A$ and the NLO low-energy constant $c_E$. The flavor factor $k_f$ depends on the considered 
channel, i.e.\ on the flavors of $B$, $B^*$ and the axial-vector current. Correspondingly, $m_{\rm GB}$ denotes the mass of the 
Goldstone-boson that can be excited in the considered channel. The flavor factor is extracted from 
$\epsilon^{ade} \, \bar T^\nu_{abc} \, (a^\mu)^b_d \, B^c_e$. 

We deduce from \eqref{eq:resHi} the following information: 
In the NLO approximation, one needs for all the axial-vector decuplet-to-octet transitions 
only two flavor symmetric low-energy constants. It does not matter in which flavor channels one determines these constants. 
The main part of this paper is devoted to suggestions how to pin down the NLO low-energy constant $c_E$. 

Obviously, $c_E$ contributes differently than $h_A$. Thus a minimal and complete NLO Lagrangian must contain a $c_E$-type 
structure. It is missing, for instance, in \cite{Jiang:2018mzd}. It might be interesting to explain why this low-energy constant
has an index ``$E$'' for ``electric''. At low energies, i.e.\ in the non-relativistic limit for the baryons, the dominant 
contribution for the $H_2$ form factor stems from $\gamma^0$ and from spatial $\nu$. This selects the combination 
$q^j a^0 - q^0 a^j$, which constitutes an electric axial-vector field strength. 

We close the present section by determining $h_A$. The corresponding structure of the LO Lagrangian gives rise to decays of 
decuplet baryons to pions and octet baryons. From each of the measured decay widths, one can extract an estimate for $h_A$. This is provided in table \ref{tab:determination_hA}, which is in agreement with \cite{Granados:2017cib}.
Note that this is an approximation for the decay widths that is accurate up to and including NLO --- because there are no 
additional contributions at NLO \cite{Holmberg:2018dtv}. 
One cannot expect to obtain always the very same numerical result for $h_A$. But the spread in the obtained values can be 
regarded as an estimate for the neglected contributions that appear beyond NLO. Indeed, those contributions break the flavor 
symmetry.

\begin{table}[H]
\centering
\begin{tabular}{|l|c|}
\hline
Decay & $h_A$ \\ \hline
$\Delta \to N \pi$   & $2.87\pm 0.05$  \\
$\Sigma^{*+} \to \Lambda \pi^+ $   & $2.39\pm0.03$  \\
$\Sigma^{*+} \to (\Sigma \pi)^+$  & $2.2\pm 0.1$  \\
$\Xi^{*0} \to (\Xi \pi)^0$        & $2.00\pm0.06$  \\ \hline
\end{tabular}
\caption{Determination of $h_A$ from different decay channels of decuplet baryons. We used mixed states in the case of the $\Delta \to N \pi$ decay. The errors come from the experimental uncertainties of the particle masses, decay widths, and branching ratios.}
\label{tab:determination_hA}
\end{table}

In section \ref{sec:chiPT} we present the basics of the relevant LO and NLO Lagrangian of baryon octet plus decuplet and Goldstone-boson octet $\chi$PT. We also specify the numerical values of the previously determined LECs. This is the first part of the paper. In section \ref{sec:Omega}, we specify the relevant interaction terms of the $\Omega^- \to \Xi^0 \, e^- \bar\nu_e$ decay and we also present the LO and NLO decay width predictions (as a function of $c_E$). Section \ref{sec:pigam} is outlined similarly to section \ref{sec:Omega} but in the case of $B(J=3/2) \to B(J=1/2) \gamma \pi$ processes. We close by showing the $c_E$ dependence of the single differential decay width of $\Xi^{*0} \to \Xi^- \pi^+ \gamma$ and finally give some concluding remarks.

\section{Chiral Lagrangian}
\label{sec:chiPT}

The relevant part of the LO chiral Lagrangian for baryons including the spin-3/2 decuplet states 
is given by \cite{Jenkins:1991es,Lutz:2001yb,Semke:2005sn,Pascalutsa:2006up,Ledwig:2014rfa, Holmberg:2018dtv} 
\begin{eqnarray}
  && {\cal L}_{\rm baryon}^{(1)} = {\rm tr}\left(\bar B \, i \slashed{D} \, B \right) + \bar T_{abc}^\mu \, i \gamma_{\mu\nu\alpha} D^\alpha (T^\nu)^{abc}
  \nonumber \\ 
  && {}+ \frac{D}{2} \, {\rm tr}(\bar B \, \gamma^\mu \, \gamma_5 \, \{u_\mu,B\}) 
  + \frac{F}{2} \, {\rm tr}(\bar B \, \gamma^\mu \, \gamma_5 \, [u_\mu,B])  \nonumber \\
  && {} + \frac{h_A}{2\sqrt{2}} \, 
  \left(\epsilon^{ade} \, \bar T^\mu_{abc} \, (u_\mu)^b_d \, B^c_e
    + \epsilon_{ade} \, \bar B^e_c \, (u^\mu)^d_b \, T_\mu^{abc} \right) \nonumber \\
  && {} -\frac{H_A}{2} \, \bar T^\mu_{abc} \gamma_\nu \gamma_5 \, (u^\nu)^c_d \; T_\mu^{abd} 
  \label{eq:baryonlagr}
\end{eqnarray}
with tr denoting a flavor trace. For mesons the well know LO chiral Lagrangian is given by \cite{Weinberg:1978kz, Gasser:1984gg, Scherer:2012xha}
\begin{eqnarray}
\label{eq:mesonlagr}
	{\cal L}_{\rm meson}^{(1)} = \frac{F_\pi^2}{4}{\rm tr}(u_\mu u^\mu)\,.
\end{eqnarray}
In principle, mass terms for the baryons mesons should be added to \eqref{eq:baryonlagr} and \eqref{eq:mesonlagr} and to the NLO Lagrangian given later. We later use the physical masses for each particle. This is a proper procedure in NLO accuracy \cite{Holmberg:2018dtv}.

We have introduced the totally antisymmetrized products of two and three 
gamma matrices \cite{pesschr},
\begin{eqnarray}
  \label{eq:defgammunu}
  \gamma^{\mu\nu} := \frac12 [\gamma^\mu,\gamma^\nu] = -i \sigma^{\mu\nu}
\end{eqnarray}
and 
\begin{eqnarray}
  \label{eq:defgammunual}
  \gamma^{\mu\nu\alpha}&:=& \frac16 
  \left(\gamma^\mu \gamma^\nu \gamma^\alpha + \gamma^\nu \gamma^\alpha \gamma^\mu + \gamma^\alpha \gamma^\mu \gamma^\nu
  \right.  \nonumber \\ && \left. \phantom{m} {}
    - \gamma^\mu \gamma^\alpha \gamma^\nu - \gamma^\alpha \gamma^\nu \gamma^\mu - \gamma^\nu \gamma^\mu \gamma^\alpha \right)
  \nonumber \\ 
  & = & \frac12 \{\gamma^{\mu\nu},\gamma^\alpha\} = +i\epsilon^{\mu\nu\alpha\beta} \gamma_\beta \gamma_5  \,,
\end{eqnarray}
respectively.
Our conventions are: $\gamma_5:=i \gamma^0 \gamma^1 \gamma^2 \gamma^3$ and 
$\epsilon_{0123}=-1$ (the latter in agreement with \cite{pesschr} 
but opposite to \cite{Pascalutsa:2006up,Ledwig:2011cx}). If a formal manipulation program is used to calculate spinor traces and 
Lorentz contractions a good check for the convention for the Levi-Civita symbol is the last relation in (\ref{eq:defgammunual}).

The spin-1/2 octet baryons are collected in ($B^a_b$ is the entry in the $a$th row, $b$th column)
\begin{equation}
  \label{eq:baroct}
  B = \left(
    \begin{array}{ccc}
      \frac{1}{\sqrt{2}}\, \Sigma^0 +\frac{1}{\sqrt{6}}\, \Lambda 
      & \Sigma^+ & p \\
      \Sigma^- & -\frac{1}{\sqrt{2}}\,\Sigma^0+\frac{1}{\sqrt{6}}\, \Lambda
      & n \\
      \Xi^- & \Xi^0 
      & -\frac{2}{\sqrt{6}}\, \Lambda
    \end{array}   
  \right)  \,.
\end{equation}
The decuplet is expressed by a totally symmetric flavor tensor $T^{abc}$ 
with 
\begin{eqnarray}
  && T^{111} = \Delta^{++} , \quad T^{112} = \frac{1}{\sqrt{3}} \, \Delta^+  , \nonumber \\
  && T^{122} = \frac{1}{\sqrt{3}} \, \Delta^0  , \quad T^{222} = \Delta^- ,    \nonumber \\
  && T^{113} = \frac{1}{\sqrt{3}} \, \Sigma^{*+}  , \quad T^{123} = \frac{1}{\sqrt{6}} \, \Sigma^{*0}  , \quad 
  T^{223} = \frac{1}{\sqrt{3}} \, \Sigma^{*-}  ,  \nonumber \\
  && T^{133} = \frac{1}{\sqrt{3}} \, \Xi^{*0} , \quad T^{233} = \frac{1}{\sqrt{3}} \, \Xi^{*-} , \quad 
  T^{333} = \Omega \,. 
  \label{eq:tensorT}
\end{eqnarray}
The Goldstone bosons are encoded in
\begin{eqnarray}
  \Phi &=&  \left(
    \begin{array}{ccc}
      \pi^0 +\frac{1}{\sqrt{3}}\, \eta 
      & \sqrt{2}\, \pi^+ & \sqrt{2} \, K^+ \\
      \sqrt{2}\, \pi^- & -\pi^0+\frac{1}{\sqrt{3}}\, \eta
      & \sqrt{2} \, K^0 \\
      \sqrt{2}\, K^- & \sqrt{2} \, {\bar{K}}^0 
      & -\frac{2}{\sqrt{3}}\, \eta
    \end{array}   
  \right) 
  \,, \nonumber \\
  u^2 & := & U := \exp(i\Phi/F_\pi) \,, \quad u_\mu := i \, u^\dagger \, (\nabla_\mu U) \, u^\dagger = u_\mu^\dagger \,. \phantom{mm}
  \label{eq:gold}
\end{eqnarray}

The fields have the following transformation properties with respect to chiral 
transformations \cite{Jenkins:1991es,Scherer:2012xha}
\begin{eqnarray}
  U \to L \, U \, R^\dagger \,, &&  u \to L \, u \, h^\dagger = h \, u \, R^\dagger  \,, \nonumber \\
  \label{eq:chiraltrafos}  
  u_\mu \to h \, u_\mu \, h^\dagger \,, &&  B \to h \, B \, h^\dagger \,,   \\
  T^{abc}_\mu \to h^a_{d} \, h^b_{e} \, h^c_{f} \, T^{def}_\mu \,,  && 
  \bar T_{abc}^\mu \to (h^\dagger)_a^{d} \, (h^\dagger)_b^{e} \, (h^\dagger)_c^{f} \, \bar T_{def}^\mu \,.  \nonumber 
\end{eqnarray}
In particular, the choice of upper and lower flavor indices is used to indicate that upper indices transform with $h$ 
under flavor transformations while the lower components transform with $h^\dagger$. 

The chirally covariant derivative for a (baryon) octet is defined by
\begin{eqnarray}
  \label{eq:devder}
  D^\mu B := \partial^\mu B + [\Gamma^\mu,B]   \,,
\end{eqnarray}
for a decuplet $T$ by
\begin{eqnarray}
  (D^\mu T)^{abc} &:=& \partial^\mu T^{abc} + (\Gamma^\mu)^a_{a'} T^{a' bc} + (\Gamma^\mu)^b_{b'} T^{a b' c} \nonumber \\
  && {} + (\Gamma^\mu)^c_{c'} T^{a bc'}   \,,
  \label{eq:devderdec}
\end{eqnarray}
for an anti-decuplet by 
\begin{eqnarray}
  (D^\mu \bar T)_{abc} &:=& \partial^\mu \bar T_{abc} - (\Gamma^\mu)_a^{a'} \bar T_{a' bc} - (\Gamma^\mu)_b^{b'} \bar T_{a b' c}  
  \nonumber \\
  && {} - (\Gamma^\mu)_c^{c'} \bar T_{a bc'}   \,,
  \label{eq:devderantidec}
\end{eqnarray}
and for the Goldstone boson fields by
\begin{eqnarray}
  \label{eq:devderU}
  \nabla_\mu U := \partial_\mu U -i(v_\mu + a_\mu) \, U + i U \, (v_\mu - a_\mu)
\end{eqnarray}
with
\begin{eqnarray}
  \Gamma_\mu &:=&  \frac12 \, \left(
    u^\dagger \left( \partial_\mu - i (v_\mu + a_\mu) \right) u \right. \nonumber \\
    && \phantom{m} \left. {}+
    u \left( \partial_\mu - i (v_\mu - a_\mu) \right) u^\dagger
  \right) \,,
  \label{eq:defGammamu}
\end{eqnarray}
where $v$ and $a$ denote external vector and axial-vector sources. 

Standard values for the coupling constants are $F_\pi=92.4\,$MeV, $D=0.80$, $F=0.46$. For the pion-nucleon coupling constant this implies $g_A=F+D =1.26$. In the case of $H_A$, we use estimates from large-$N_c$ considerations: $H_A = \frac95 \, g_A \approx 2.27$ \cite{Pascalutsa:2006up,Ledwig:2011cx} or $H_A = 9F -3D \approx 1.74$ \cite{Dashen:1993as,Semke:2005sn}, since we lack a simple direct observable to pin it down. Numerically we use $H_A \approx 2.0$. 

At NLO, we have five terms that contribute to the decays of interest \cite{Holmberg:2018dtv}: two octet sector terms fields that are given by
\begin{equation}
 \label{eq:NLO1}
 	b_{M,D} \, {\rm tr}(\bar{B}\{ f_+^{\mu\nu},\sigma_{\mu\nu} B\}) + b_{M,F} \, {\rm tr}(\bar{B}[ f_+^{\mu\nu},\sigma_{\mu\nu} B])\,,
\end{equation}
one term from the decuplet sector given by
\begin{eqnarray}
 \label{eq:NLO2}
	d_M \, i \, (\bar T_\mu)_{abc} \, (f^{\mu\nu}_+)^c_d \, T_\nu^{abd}\,,
\end{eqnarray}
and finally two decuplet-to-octet transition terms given by
\begin{eqnarray}
  &i c_M \epsilon_{ade} \, \bar B^e_c \, \gamma_\mu \gamma_5 (f_+^{\mu\nu})^d_b \, T_\nu^{abc} \nonumber \\ [0.3em]
  & {} + i \, c_E \epsilon_{ade} \, \bar B^e_c \, \gamma_\mu (f_-^{\mu\nu})^d_b \, T_\nu^{abc} + \text{h.c.} \,.
  \label{eq:NLO3}
\end{eqnarray}
The field strengths $f_\pm^{\mu\nu}$ are given by
\begin{eqnarray}
  \label{eq:deffpm}
  f_\pm^{\mu\nu} := u \, F_L^{\mu\nu} \, u^\dagger \pm u^\dagger \, F_R^{\mu\nu} \, u
\end{eqnarray}
with
\begin{equation}
  F_{R,L}^{\mu\nu} := \partial^\mu \, (v^\nu \pm a^\nu) - \partial^\nu \, (v^\mu \pm a^\mu) -i \, [v^\mu \pm a^\mu,v^\nu \pm a^\nu]  \,.
  \label{eq:defFRL}
\end{equation}
Interactions with external forces are studied by the replacement of the vector and axial-vector sources $v^\nu$, $a^\nu$. For electromagnetic interactions we have the replacement \cite{Scherer:2002tk}
\begin{eqnarray}
  \label{eq:emvA}
  v_\mu \to e A_\mu \left(
    \begin{array}{rrr}
      \frac23 & 0 & 0  \\[0.5em]
      0 & -\frac13 & 0 \\[0.5em]
      0 & 0 & -\frac13
    \end{array}
    \right)
\end{eqnarray}
with the photon field $A_\mu$ and the proton charge $e$; and for weak interactions mediated by the W-bosons we have the replacement \cite{Scherer:2002tk}
\begin{eqnarray} \label{eq:l}
	v_\mu - a_\mu \to -\frac{g_w}{\sqrt{2}}W_\mu^+ \left( {\begin{array}{*{20}{c}}
0&V_{ud}&V_{us}\\
0&0&0\\
0&0&0
\end{array}} \right) + \text{h.c.}
\end{eqnarray}
with the W-boson field $W_\mu^+$, the Cabibbo-Kobayashi-Maskawa matrix elements $V_{ud}$, $V_{us}$ \cite{Kobayashi:1973}, and the weak gauge coupling $g_w$ (related to Fermi's constant and the W mass). 

The values of the low-energy constants $b_{M,D/F}$ and $d_M$ are determined by fitting the calculated and measured magnetic moments of the octet and decuplet baryons, respectively. The results are: $ b_{M,D} \approx 0.321 \, \mbox{GeV}^{-1}$, $b_{M,F} \approx 0.125 \, \mbox{GeV}^{-1}$, and $d_M \approx -1.16 \,\mbox{GeV}^{-1}$. Furthermore, from the radiative decay of decuplet baryons to a pion and a photon, one obtains $c_M \approx \pm 1.92 \, \mbox{GeV}^{-1}$. The values of the above NLO LECs come from \cite{Holmberg:2018dtv}. Next we present two types of decay channels that can be used to determine $c_E$ and the sign of $c_M$, starting with $\Omega^- \to \Xi^0 \, e^- \bar\nu_e$.

In view of the absence of data that one can use to help pin down $c_E$ and the sign of $c_M$; we stress that the purpose of this paper is to motivate such experiments and not to explore the uncertainties of already determined LECs. Therefore, we stick to the central values of those LECs for numerical results.

\section{Decay process $\Omega^- \to \Xi^0 \, e^- \bar\nu_e$}
\label{sec:Omega}

\begin{figure}[H] 
\centering
  \begin{subfigure}[b]{0.45\linewidth}
    \centering
    \includegraphics[width=1\linewidth]{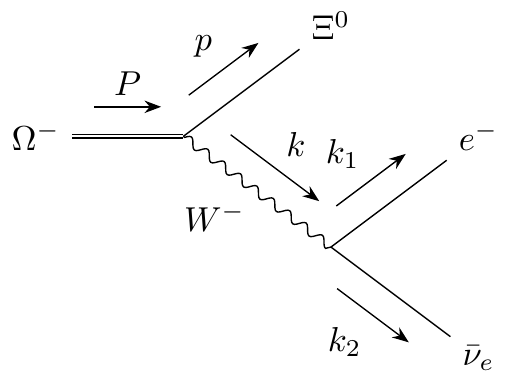} 
  \end{subfigure}
  \begin{subfigure}[b]{0.55\linewidth}
    \centering
    \includegraphics[width=1\linewidth]{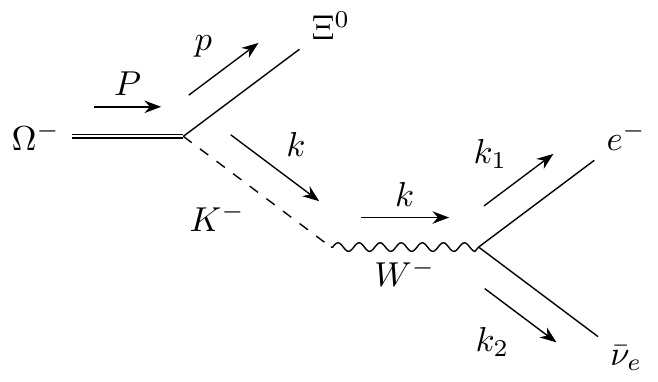} 
  \end{subfigure} 
  \caption{Diagrams contributing to the decay of $\Omega^- \to \Xi^0 \, e^- \bar\nu_e$. These two diagrams are the only two topologically distinct diagrams at NLO, but note that the $\Omega\Xi W$-vertex comes from three terms.}
  \label{fig:Omega_diagrams}
\end{figure}

There are two contributing tree-level diagrams at NLO in $\chi$PT for the decay of the Omega baryon decaying into a Cascade baryon, an electron, and an electron anti-neutrino. These are shown in figure \ref{fig:Omega_diagrams}.

\begin{figure}[H] 
    \centering
    \includegraphics[width=0.9\linewidth]{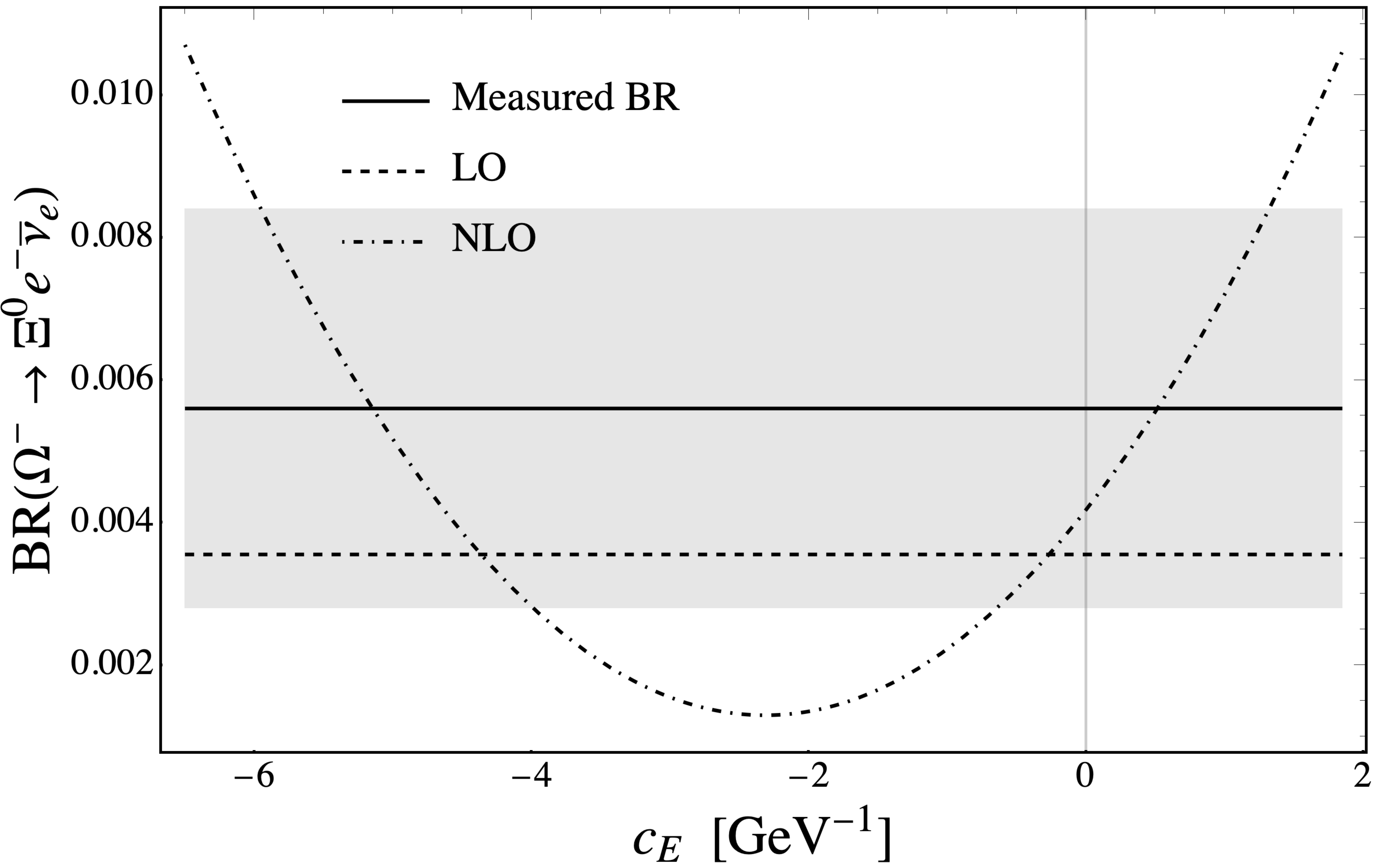} 
    \caption{NLO branching ratio of $\Omega^- \to \Xi^0 e^- \bar{\nu}_e$ as a function of $c_E$, together with the LO result and the experimental value \cite{pdg}. The gray box illustrates the measurement uncertainty.}
    \label{fig:omega_decay_cE}
\end{figure}

\noindent At LO, we have three contributing terms: the $\sim h_A$ term from \eqref{eq:baryonlagr}, the kinetic term for the mesons in \eqref{eq:mesonlagr}, as well as the standard weak charged-current interaction term \cite{Scherer:2002tk}. Going to NLO, we also have the $\sim c_{M/E}$ terms in \eqref{eq:NLO3}. When calculating the decay width we used Mathematica and FeynCalc to perform the traces of the gamma matrices \cite{Mathematica, FeynCalc}. The resulting partial decay width at NLO contains terms proportional to $h_A^2$, $c_E^2$, $c_M^2$ and $h_A c_E$. Figure \ref{fig:omega_decay_cE} illustrates the calculated and measured branching ratio. We use the values of the LECs described in section \ref{sec:chiPT} together with $h_A = 2.0$, i.e., the value of $h_A$ obtained from fitting the partial decay width of $\Xi^* \to \Xi \pi$ to data, again see table \ref{tab:determination_hA}. 

\vspace{1mm}

We then fit the NLO branching ratio to the measurement, resulting in $c_E$: $(0.5 \pm 1)$ GeV$^{-1}$ and $(-5 \pm 1)$ GeV$^{-1}$. The error comes from the experimental uncertainty of the branching ratio. We cannot distinguish between the two solutions of $c_E$ by only considering the integrated decay width, and likewise, we cannot pin down the sign of $c_M$ since the partial decay width contains no linear $c_M$ term.

\begin{figure}[H] 
  \begin{subfigure}[b]{0.53\linewidth}
    \centering
    \includegraphics[width=0.9\linewidth]{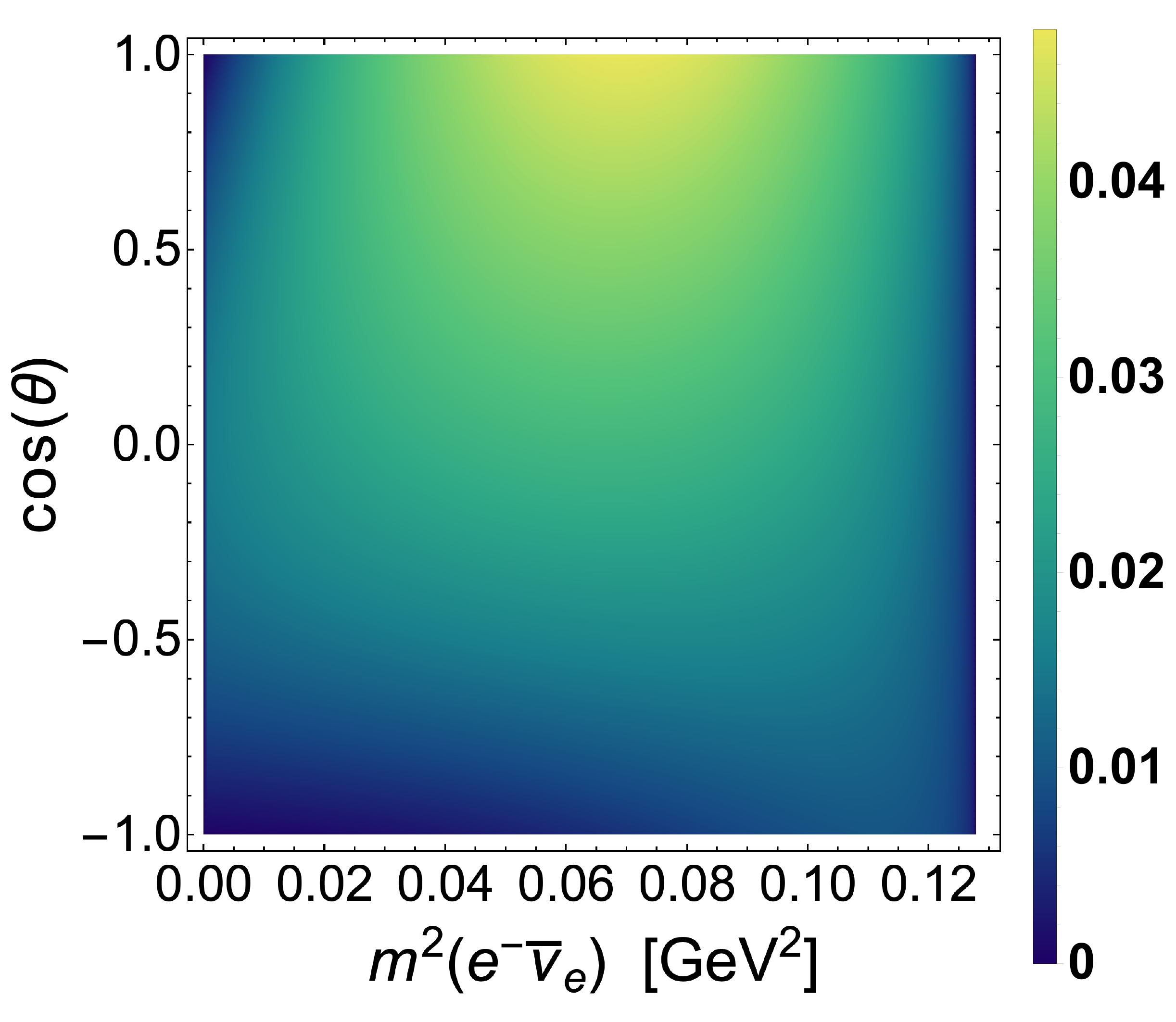} 
    \caption{$c_M = 1.92$ GeV$^{-1}$, \\$c_E = 0.5$ GeV$^{-1}$} 
  \end{subfigure}
  \begin{subfigure}[b]{0.53\linewidth}
    \centering
    \includegraphics[width=0.91\linewidth]{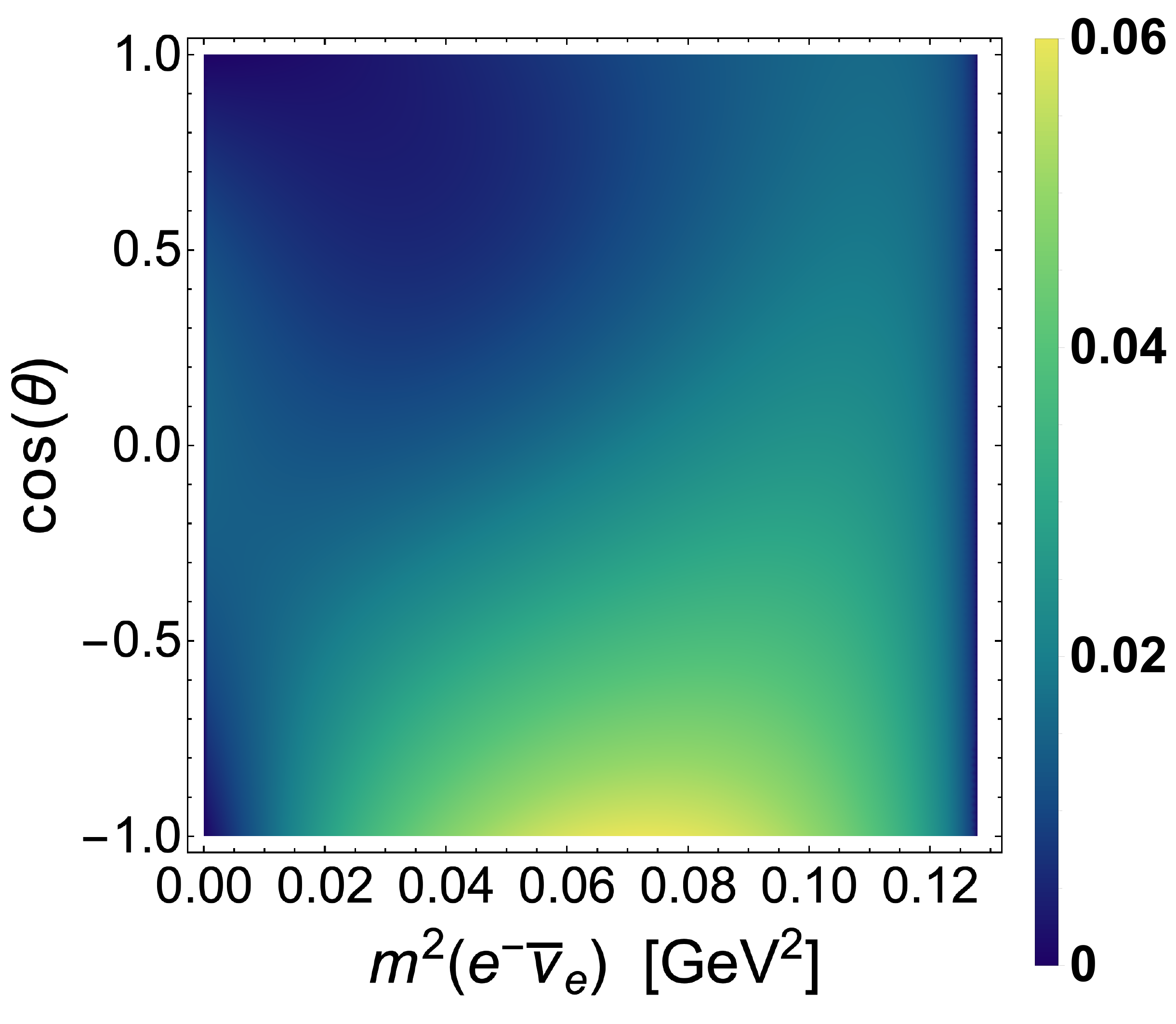} 
    \caption{$c_M = 1.92$ GeV$^{-1}$, \\$c_E = -5$ GeV$^{-1}$} 
  \end{subfigure} 
  \begin{subfigure}[b]{0.53\linewidth}
    \centering
    \includegraphics[width=0.9\linewidth]{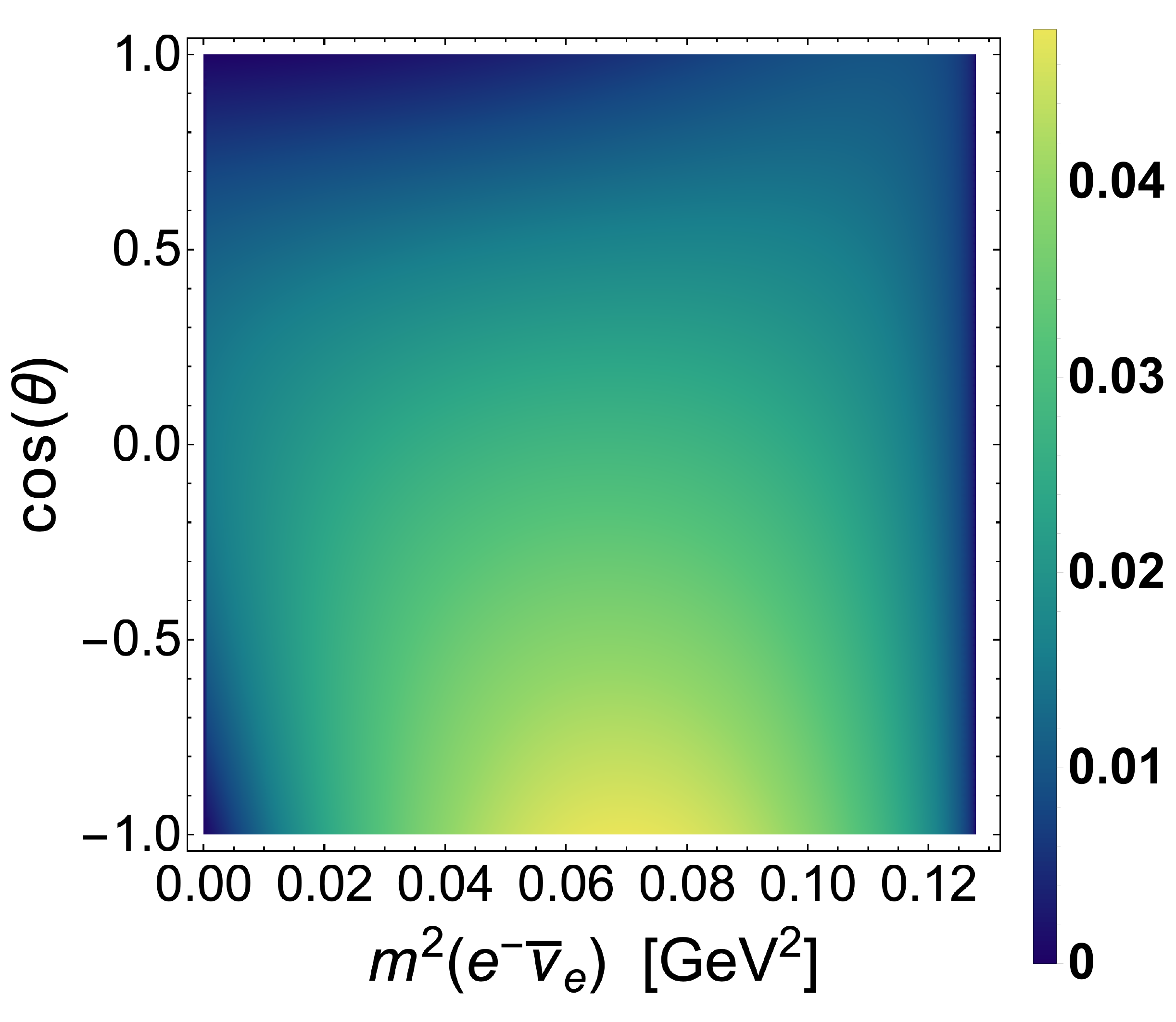} 
    \caption{$c_M = -1.92$ GeV$^{-1}$, \\$c_E = 0.5$ GeV$^{-1}$} 
  \end{subfigure}
  \begin{subfigure}[b]{0.53\linewidth}
    \centering
    \includegraphics[width=0.91\linewidth]{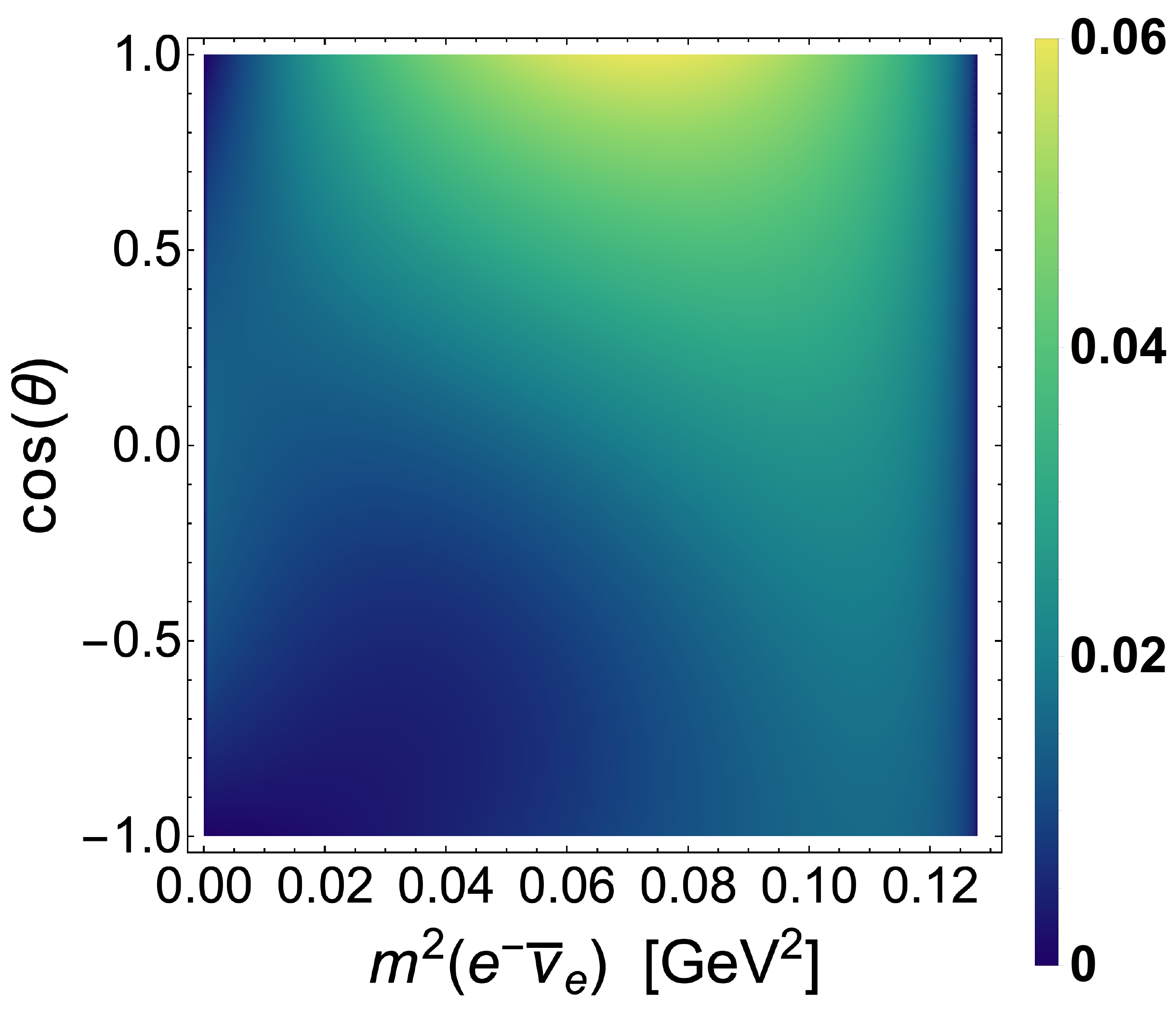} 
    \caption{$c_M = -1.92$ GeV$^{-1}$, \\$c_E = -5$ GeV$^{-1}$} 
  \end{subfigure} 
  \caption{The double differential decay distribution of $\Omega^- \to \Xi^0 \bar{\nu}_e e^-$ for the two solutions of $c_E$ and for different signs of $c_M$. The scale of the double differential decay distribution is linear with an arbitrary normalization.} 
  \label{fig:omega_double_decay_rate} 
\end{figure}

We can instead study the distribution of the double differential decay width to obtain more information. In figure \ref{fig:omega_double_decay_rate} we illustrate the double differential decay distribution for the four different solutions of $c_M$ and $c_E$ in the frame where the electron and anti-neutrino goes back to back, i.e., $\vec k_1 + \vec k_2 = 0$. We chose to work with the kinematic variables $m^2(e^-\bar \nu_e)$ and $\cos(\theta)$, where $\theta$ is the angle between the three-momenta of the Cascade baryon and the electron. With enough statistics, it would be possible to distinguish between all four cases, and even with less statistics, it could be possible to at least determine the relative sign of $c_M$ and $c_E$, depending on where one finds the majority of events in the Dalitz plots (i.e., closer to $\cos(\theta) = 1$ or $\cos(\theta) = -1$). Note also that there is an apparent antisymmetry such that $c_M \to -c_M$ is equivalent to $\cos(\theta) \to - \cos(\theta)$. This is, however, only approximately true when the lepton masses are small compared to the hadron masses. In the case of vanishing lepton masses, we find that all linear terms of $\cos(\theta)$ in the double differential decay width are proportional to $c_M$. This phenomenon is called forward-backward asymmetry \cite{donoghue1994dynamics}.

\section{Decay processes $B(J=3/2) \to B(J=1/2) \gamma \pi$}
\label{sec:pigam}

\begin{figure}[H] 
\centering
  \begin{subfigure}{0.29\linewidth}
    \centering
    \includegraphics[width=1\linewidth]{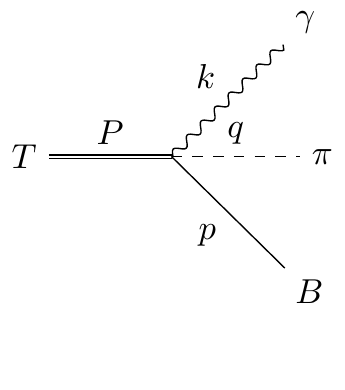}
    \caption*{1.} 
  \end{subfigure}\quad
  \begin{subfigure}{0.4\linewidth}
    \centering
    \includegraphics[width=1\linewidth]{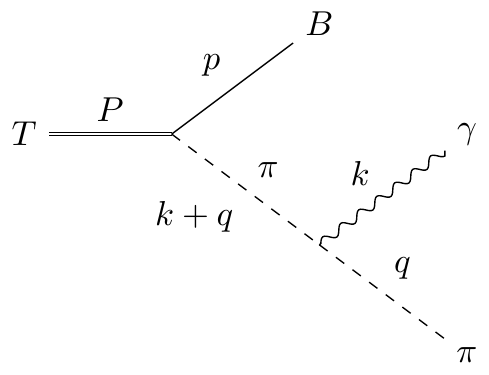} 
    \caption*{2.} 
  \end{subfigure} \par
  \begin{subfigure}{0.4\linewidth}
    \centering
    \includegraphics[width=1\linewidth]{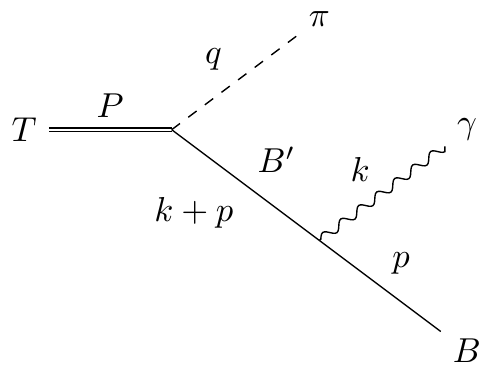} 
    \caption*{3.} 
  \end{subfigure} \quad
  \begin{subfigure}{0.4\linewidth}
    \centering
    \includegraphics[width=1\linewidth]{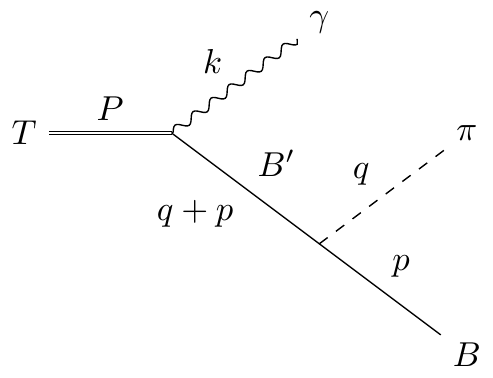} 
    \caption*{4.} 
  \end{subfigure} \par
  \begin{subfigure}{0.4\linewidth}
    \centering
    \includegraphics[width=1\linewidth]{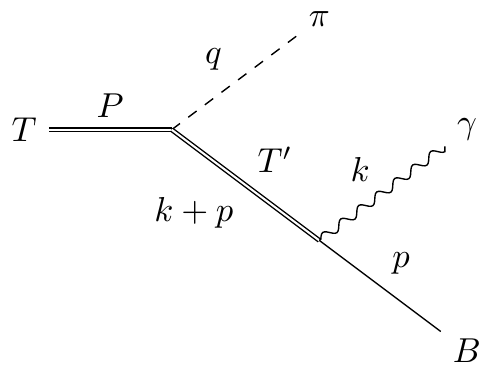} 
    \caption*{5.} 
  \end{subfigure} \quad
  \begin{subfigure}{0.4\linewidth}
    \centering
    \includegraphics[width=1\linewidth]{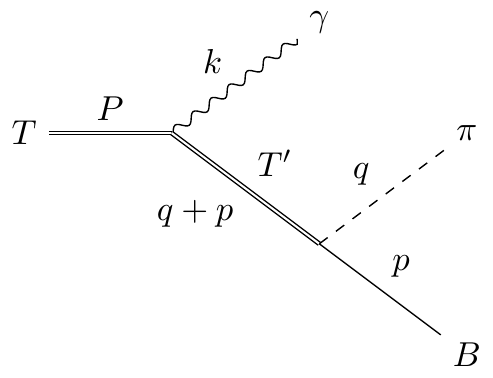} 
    \caption*{6.} 
  \end{subfigure}\par 
  \caption{Diagrams contributing to the decay of a decuplet baryon into an octet baryon, a pion and a photon.}
  \label{fig:Radiative_threebody_diagrams}
\end{figure}

For the decay of a decuplet baryon into an octet baryon, a pion and a photon we have 6 diagrams at NLO in $\chi$PT. These diagrams are shown in figure \ref{fig:Radiative_threebody_diagrams}. The interaction terms are given by all terms in \eqref{eq:baryonlagr}, \eqref{eq:mesonlagr}, \eqref{eq:NLO1}, \eqref{eq:NLO2}, and \eqref{eq:NLO3}.

As before, we use Mathematica and FeynCalc to perform the traces of the gamma matrices. Furthermore, we explicitly checked that the Ward identity for the electromagnetic current holds, that is $\mathcal{M}_\mu k^\mu = 0$ for all decays.

Concerning the values of the different LECs, we used $h_A = 2.4$, being the average value of table \ref{tab:determination_hA}, together with $c_M = + 1.92 \text{ GeV}^{-1}$ and $c_E = 0.5 \text{ GeV}^{-1}$. The decays contain infrared divergences resulting from final-state radiation that, in the limit of vanishing photon energy, can make a propagating particle on-shell \cite{pesschr}. Therefore, we used a cutoff of the photon energy at $25$ MeV (in the frame where $\vec p + \vec q = 0$), roughly corresponding to the lowest detectable photon energy at the upcoming $\bar{\mbox{P}}$ANDA experiment (Antiproton ANnihilation at DArmstadt) \cite{Thome:2012bdy, Lutz:2009ff, Erni:2008uqa} and the Beijing Spectrometer III \cite{Ablikim:2009aa}. This cutoff can be arbitrarily chosen to match the photon energy resolution of any experiment. In table \ref{tab:Three_body_BR}, we have collected the predictions of all energetically possible decays and their branching ratios at LO and NLO as well as the LEC dependence.

Looking at table \ref{tab:Three_body_BR} we see that decay widths containing only neutral states vanish at LO, which is only natural since neutral hadrons do not interact with photons at LO. Contributions with LO LECs appear at NLO since the amplitudes contain structures which are proportional to products of LO and NLO LECs, e.g., diagram 3 gives terms like $\sim h_A b_{M,D}$, but with absent pure LO contributions. Furthermore, it is reassuring that the branching ratios of these neutral decays are small at NLO since they vanish at LO; indicating that the NLO contribution is, in general, a small correction.

The branching ratios of decays with neutral pions in the final state are small due to the same reason, that is, since the pion-pion-photon vertex forbids neutral pions, the otherwise large contribution of the pion propagator in diagram 2 disappear. 

Let us briefly investigate the possible ranges of the different LECs due to their uncertainty, starting with $H_A$. Varying the value of $H_A$ by $\pm 0.3$ changes the decay widths insignificantly (much less than 1\%), except in the cases of $\Xi^{*0}\,\to\,\Xi^0\,\pi^0\,\gamma$ and $\Delta^{0}\,\to\, n\, \pi^0\,\gamma$ which change by $10$\% and $2$\%, respectively. We also considered $c_M = - 1.92 \text{ GeV}^{-1}$, which changes the decay widths by a few percents (often much less) in the case of decays involving charged states. The four decays with only neutral states (and vanishing decay width at LO) changed significantly (up to 80\%). But since they vanish at LO they are prone to be more sensitive to the values of the NLO LECs.

\begin{figure}[H] 
    \centering
    \includegraphics[width=0.9\linewidth]{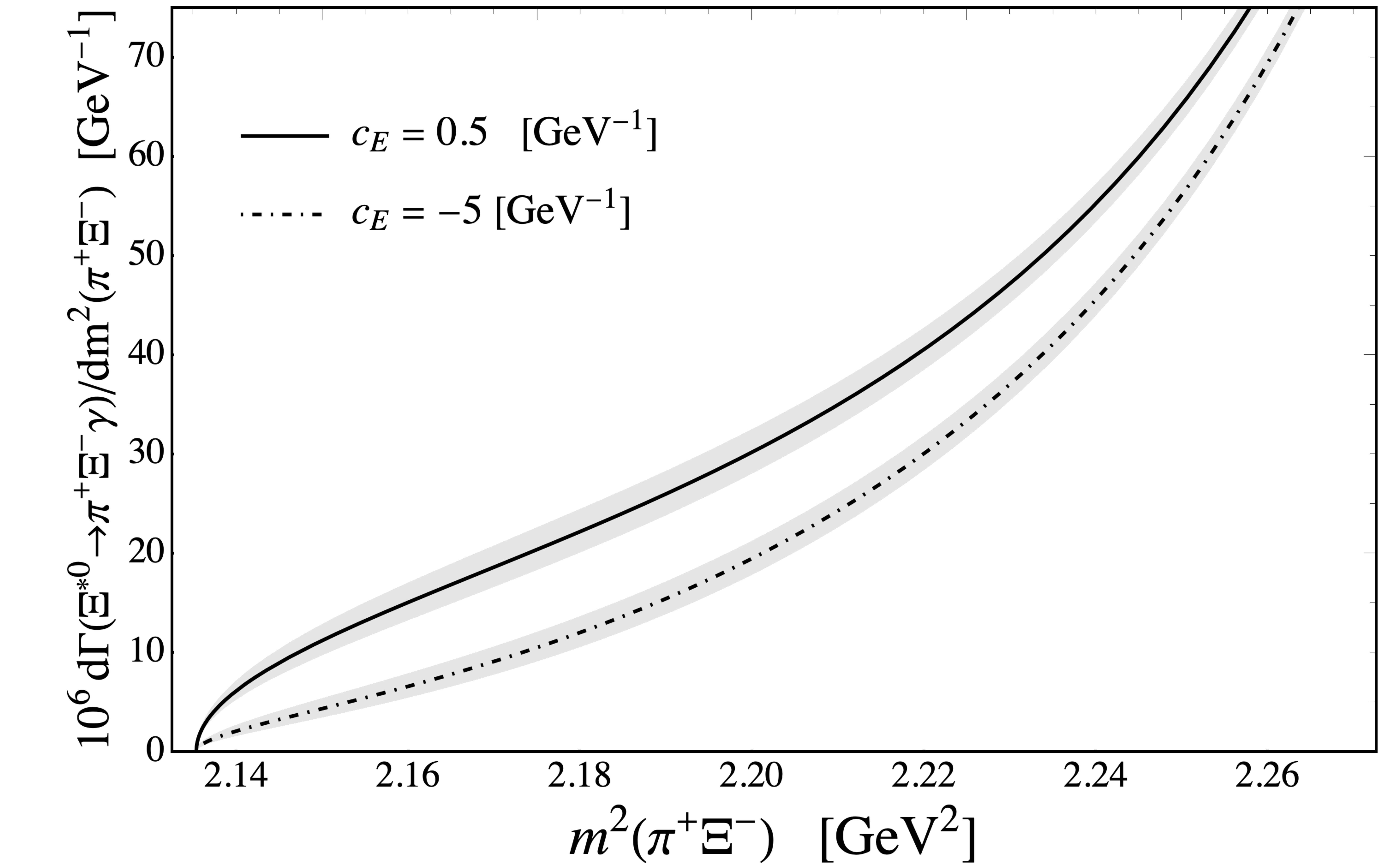} 
  \caption{Impact of the two solutions of $c_E$ on the single differential decay width of $\Xi^{*0} \to \Xi^- \pi^+ \gamma$ using $c_M = +1.92$ GeV$^{-1}$. We plot the single differential decay width down to a photon energy of $50$ MeV (otherwise the $c_E$-dependence is hard to display). The gray intervals come from the uncertainty of the two solutions of $c_E$.} 
  \label{fig:three_body_decay_single}
\end{figure}

Once data are available, we can use these radiative three-body decays to further investigate $c_E$. For this purpose, the decays $\Xi^{*-}\,\to\,\Xi^0\,\pi^-\,\gamma$ and $\Xi^{*0}\,\to\,\Xi^-\,\pi^+\,\gamma$ are of most interest, since they both have a relatively large branching ratio of $\sim 10^{-3}$ and because of the small (total) decay width of Cascade baryons (as compared to broad Delta baryons). We find that the decay widths of these two Cascade decays decrease by around 20\% when changing to the negative $c_E$ solution. In figure \ref{fig:three_body_decay_single} we consider how the single differential decay width of $\Xi^{*0}\,\to\,\Xi^-\,\pi^+\,\gamma$ changes when varying $c_E$.
\vspace{1mm}

We note that at low photon energies both solutions of $c_E$ converge, in the region where the single differential decay width blows up, due to the infrared divergence. Instead, it is at higher photon energies, i.e., lower $m^2(\pi^+ \Xi^-) = M_{\Xi}^2 - 2 M_{\Xi} E_\gamma$, where we can distinguish between the two possible $c_E$.

\begin{widetext}

\begin{table}[H]
\centering
\small
\begin{tabular}{|l|l|c|c|}
\hline
Decay  & LEC dependence at NLO & BR at LO & BR at NLO\\ \hline
$\Xi^{*-}\,\to\,\Xi^-\,\pi^0\,\gamma$ & $h_A,\,d_M,\,b_{M,D},\,b_{M,F}$ & $8.2 \times 10^{-6}$ & $8.6 \times 10^{-6}$\\
$\Xi^{*-}\,\to\,\Xi^0\,\pi^-\,\gamma$  & $h_A,\,H_A,\,c_M,\,c_E,\,d_M,\,b_{M,D}$ & $1.4 \times 10^{-3}$ & $1.4 \times 10^{-3}$\\
$\Xi^{*0}\,\to\,\Xi^-\,\pi^+\,\gamma$  & $h_A,\,D,\,F,\,c_M,\,c_E,\,b_{M,D},\,b_{M,F}$ & $1.2 \times 10^{-3}$ & $1.2 \times 10^{-3}$ \\
$\Xi^{*0}\,\to\,\Xi^0\,\pi^0\,\gamma$ & $h_A,\,H_A,\,D,\,F,\,c_M,\,b_{M,D}$ & $0$ & $1.9 \times 10^{-6}$ \\
$\Sigma^{*+}\,\to\,\Sigma^+\,\pi^0\,\gamma$ & $h_A,\,H_A,\,D,\,c_M,\,d_M,\,b_{M,D},\,b_{M,F}$ & $8.9\times 10^{-7}$ & $1.2 \times 10^{-6}$\\
$\Sigma^{*+}\,\to\,\Sigma^0\,\pi^+\,\gamma$ & $h_A,\,H_A,\,F,\,c_M,\,c_E,\,d_M,\,b_{M,D}$ & $3.6 \times 10^{-5}$ & $3.7 \times 10^{-5}$\\
$\Sigma^{*-}\,\to\,\Sigma^-\,\pi^0\,\gamma$  & $h_A,\,d_M,\,b_{M,D},\,b_{M,F}$ & $6.3 \times 10^{-7}$ & $6.6 \times 10^{-7}$  \\
$\Sigma^{*-}\,\to\,\Sigma^0\,\pi^-\,\gamma$ & $h_A,\,H_A,\,c_M,\,c_E,\,d_M,\,b_{M,D}$ & $4.4 \times 10^{-5}$ & $4.5 \times 10^{-5}$ \\
$\Sigma^{*0}\,\to\,\Sigma^+\,\pi^-\,\gamma$ & $h_A,\,H_A,\,D,\,F,\,c_M,\,c_E,\,b_{M,D},\,b_{M,F}$ & $5.9 \times 10^{-5}$ & $5.9 \times 10^{-5}$ \\
$\Sigma^{*0}\,\to\,\Sigma^-\,\pi^+\,\gamma$ & $h_A,\,D,\,F,\,c_M,\,c_E,\,b_{M,D},\,b_{M,F}$ & $3.3 \times 10^{-5}$ &  $3.4 \times 10^{-5}$\\
$\Sigma^{*0}\,\to\,\Sigma^0\,\pi^0\,\gamma$  & $h_A,\,D,\,c_M,\,b_{M,D}$ & $0$ & $2.6 \times 10^{-8}$ \\
$\Sigma^{*0}\,\to\,\Lambda\,\pi^0\,\gamma$ & $h_A,\,D,\,c_M,\,b_{M,D}$ & $0$ &  $3.6 \times 10^{-6}$ \\
$\Delta^{++}\,\to\, p\, \pi^+\,\gamma$  & $h_A,\,H_A,\,c_M,\,c_E,\,d_M,\,b_{M,D},\,b_{M,F}$ & $1.7 \times 10^{-3}$ &  $1.8 \times 10^{-3}$ \\
$\Delta^{+}\,\to\, p\, \pi^0\,\gamma$  & $h_A,\,H_A,\,D,\,F,\,c_M,\,d_M,\,b_{M,D},\,b_{M,F}$ & $5.6 \times 10^{-5}$ & $7.2 \times 10^{-5}$ \\
$\Delta^{+}\,\to\, n\, \pi^+\,\gamma$ & $h_A,\,H_A,\,D,\,F,\,c_M,\,c_E,\,d_M,\,b_{M,D}$ & $7.5 \times 10^{-4}$ & $7.6 \times 10^{-4}$ \\
$\Delta^{0}\,\to\, p\, \pi^-\,\gamma$  & $h_A,\,H_A,\,D,\,F,\,c_M,\,d_M,\,b_{M,D},\,b_{M,F}$ & $1.0 \times 10^{-3}$ & $1.0 \times 10^{-3}$ \\
$\Delta^{0}\,\to\, n\, \pi^0\,\gamma$  & $h_A,\,H_A,\,D,\,F,\,c_M,\,b_{M,D}$ & $0$ & $7.6 \times 10^{-6}$ \\
$\Delta^{-}\,\to\, n\, \pi^-\,\gamma$  & $h_A,\,H_A,\,c_M,\,c_E,\,d_M,\,b_{M,D}$& $2.3 \times 10^{-3}$ & $2.3 \times 10^{-3}$  \\ \hline                    
\end{tabular}
\caption{Branching ratios at LO and NLO of all energetically possible $B^*(J=3/2) \to B \pi \gamma$ decays allowed by flavor symmetry, except $\Sigma^{*\pm}\,\to\,\Lambda\,\pi^\pm\,\gamma$ that run over the pole of $\Sigma^0$. The second column is showing the LEC dependence at NLO.} 
\label{tab:Three_body_BR}
\end{table}

\end{widetext}

To conclude, we provide two types of decay channels that can be used to probe the axial-vector transition of a decuplet to an octet baryon, parameterized by $c_E$. The most promising decays for this purpose are $\Omega^- \to \Xi^0 \, e^- \bar\nu_e$, $\Xi^{*-}\,\to\,\Xi^0\,\pi^-\,\gamma$, and $\Xi^{*0}\,\to\,\Xi^-\,\pi^+\,\gamma$. Moreover, by studying the double differentiable decay distribution of the Omega decay, one can determine the sign of $c_M$. We hope that such experiments can be carried out at $\bar{\mbox{P}}$ANDA and in part at BES III.

\acknowledgments{The authors thank Albrecht Gillitzer for valuable discussions and his interest in a future measurement of the radiative three-body decays using the $\bar{\mbox{P}}$ANDA experiment.}


\bibliography{lit}{}
\bibliographystyle{apsrev4-1}
\end{document}